\documentclass[onecollarge,natbib]{svjour2}
\bibpunct{[}{]}{;}{n}{}{,} 
\smartqed  
\usepackage{graphicx}
\usepackage{comment}
\usepackage{amsmath,amssymb,amsfonts,textcomp}

\def\eg{{\it e.g.}}
\journalname{Few-Body Systems}
\begin{document}

\title{Double parton distributions in Light-Front constituent
quark models
}


\author{Matteo Rinaldi        \and
        Sergio Scopetta \and Marco Traini \and Vicente Vento 
}


\institute{M. ~Rinaldi \at
              Dipartimento di Fisica e Geologia, Universit\`a degli Studi di
              Perugia, and INFN, sezione di Perugia, 
              06100 Perugia, Italy \\
              \email{matteo.rinaldi@pg.infn.it}    \\
              Tel. $+39~ 075~ 585 2793$ 
           \and
           S.~Scopetta \at
              Dipartimento di Fisica e Geologia, Universit\`a degli Studi di
              Perugia, and INFN, sezione di Perugia, 
              06100 Perugia, Italy  
                         \and
           M.~Traini \at
              Dipartimento di Fisica, Universit\`a degli studi di Trento, and
              INFN - TIFPA, 38123 Trento, Italy 
                         \and
           V.~Vento \at
              Departament de Fisica Te\`orica, Universitat de Val\`encia
and Institut de Fisica Corpuscular, Consejo Superior de Investigaciones
Cient\'{\i}ficas, 46100 Burjassot (Val\`encia), Spain 
}

\date{Received: date / Accepted: date}

\maketitle

\begin{abstract}
Double parton distribution functions (dPDF), accessible in high
energy proton-proton and proton nucleus collisions, encode information
on how partons inside a proton are correlated among each other
and could represent a tool to explore the 3D proton structure.
In recent papers, double parton correlations have been studied 
in the valence quark region, by means of
constituent quark models. 
This framework allows to understand 
clearly the dynamical origin of the correlations and to
establish which, among the features of the results, 
are model independent. 
Recent relevant  results, obtained in a relativistic 
light-front scheme, able to overcome some drawbacks of previous 
calculations, such as the poor support, 
will  be presented. Peculiar
transverse momentum correlations, 
generated by the correct treatment
of the boosts, are obtained. The role  of spin  correlations will be 
also shown. 
In this covariant approach, the symmetries of the dPDFs
are unambiguously reproduced. The study of the QCD 
evolution of the model results has been
performed in the valence sector, showing that, in some cases,
the effect of evolution does not cancel that of correlations.
\end{abstract}

\section{Introduction}
\label{intro}
In high energy hadron-hadron collisions,
more than one parton in each of the
hadron can contribute to the cross section. This is the
multiple partonic interactions (MPI) phenomenon which,
even if its contribution is suppressed by a power of
$\Lambda^2_{QCD}/Q^2$ with respect to the single parton interaction, with
$Q$ the center-mass energy in the collision, has 
been already observed (see, \eg, Ref. \cite{2a}). In this scenario MPI
represent a background for the search of new  Physics, \eg, at the LHC.
In this work we focus our
attention to the double parton scattering (DPS) which can be
observed in many channels, \eg, $WW$ with dilepton productions
and double
Drell-Yan processes (see, {\it e.g.} Refs. \cite{3a,4a,5a,6a} for recent 
reviews).  
At the LHC, DPS, whose evidence 
has been observed \cite{16a}, represents also a
background for Higgs production.  In this framework the DPS cross section
can be written, following \cite{1a}, in terms of the so
called double parton distribution functions (dPDFs),
$F_{ij}(x_1,x_2,{\vec z}_\perp,\mu)$, which describe the 
 joint probability 
of finding two partons of flavors $i,j=q, \bar q,g$ with 
longitudinal momentum fractions $x_1,x_2$ and transverse separation 
$\vec z_\perp$ inside the hadron:
\begin{eqnarray}  \label{eq:si_old}
  d \sigma  = \frac{1}{S} \sum_{i,j,k,l} \int\! d \vec z_\perp\, 
F_{ij}(x_1,x_2,\vec z_\perp,\mu) 
F_{kl}(x_3,x_4,\vec z_\perp,\mu) 
   \hat \sigma_{ik}(x_1 x_3 \sqrt{s},\mu) \hat \sigma_{jl}(x_2 x_4
\sqrt{s},\mu)
\,.\end{eqnarray}

The partonic cross sections $\hat \sigma$ refer to the hard,
short-distance processes,
$S$ is a symmetry factor, present if identical particles appear in 
the final state and $\mu$ is the renormalization scale which is taken, for
simplicity, to be 
the same for both partons.
For the evaluation of
the DPS contributions
to proton-proton scattering at LHC kinematics,
the following approximation,
for the dPDF,
is usually made: 

\begin{equation} 
\label{app}
  F_{ij}(x_1,x_2,\vec z_\perp,\mu) = q_{i}(x_1,\mu)
q_{j}(x_2,\mu)\, \theta(1-x_1-x_2) (1-x_1-x_2)^n
T(\vec z_\perp,\mu)~,
\end{equation}
{\it i.e.},
a complete factorized form of the dPDF is assumed. In
particular the ${\bf z_\perp}$ and $x_1-x_2$ dependences 
are factorized and
the standard single parton distribution functions (PDF), $q(x)$,
are introduced. This means that possible double parton
correlations between the two interacting partons are neglected.
dPDFs are non
perturbative quantities so that they cannot be easily evaluated in
QCD. As it happens for the PDF, a useful procedure for their estimate
is a calculation at the
hadronic scale, $Q_0 \sim \Lambda_{QCD}$, by means of quark models.
In order to compare the obtained results with data taken, \eg, at high
energy scales, $Q > Q_0$, it is necessary a
perturbative QCD (pQCD) evolution of the model calculations, 
using the dPDFs evolution equations known since a long time ago
\cite{23a,24a}.
By using this procedure, future data
analysis of the DPS processes could be guided, in principle, by 
model calculations. The first model evaluations of the dPDF have been the ones
in Refs.\cite{33a,36a}. In the first scenario use has been made of a
modified version of the MIT bag model in the cavity-approximation in order
to introduce  double parton correlations by hand, recovering 
momentum conservation. In the second case the dPDFs have been calculated in
a non relativistic (NR) constituent quark model (CQM) framework, since CQM, 
in the valence region, 
predict PDFs,
generalized parton distribution functions (GPDs) and transverse momentum
dependent parton distributions (TMDs) rather well (see, \eg,
Refs. \cite{37a,40a,41a}).
These
expectations motivated the
analysis of Refs. \cite{36a, nostro} and the present one. The main results 
found
in Refs.  \cite{33a,36a} are that, in the valence quark region, the
approximations used to write Eq. (\ref{app}) are badly violated.  
In the CQM
picture, where the dynamical
origin of double correlations is clear, the origin
of this violation can be properly understood.
One should notice that both the analyses
of Refs. \cite{33a,36a} have some inconsistencies. 
First, dPDF do not vanish in the non physical
region, $x_1+x_2 > 1$,{\it i.e.}, they
have a wrong support. Moreover, as already pointed out, in order to
obtain some information on the dPDF at small values of $x$ and
at high $Q^2$, where LHC data are taken, the pQCD
evolution of the calculated dPDF is necessary.
In a recent paper of ours \cite{nostro}, a CQM calculation of the dPDF
has been performed including relativity through
a fully Poincar\'e covariant Light-Front (LF) approach.  Thanks to
this treatment it is possible to study strong interacting systems with a
fixed number of on-shell constituents
(see Refs. \cite{46a,47a} for general reviews). { Moreover, being the
hyperplane of the LF, {\it i.e.}, the plane where the initial conditions
are defined, tangent to the Light-Cone, the
Deep Inelastic Scattering (DIS) phenomenology is automatically included into
the
scheme.}
In particular, in this framework,
which has been used extensively for hadronic calculations, (see, \eg, Refs.
\cite{48a,49a,50a}, some symmetries of the dPDF are restored and
the  bad support problem is fixed, so that
the pQCD evolution of the dDPDFs is more
precise. The results of this analysis will
be summarized in the following sections.

\section{dPDFs in light-Front CQM}

In this section the procedure adopted for the LF calculation of the
dPDFs will be presented. The validity
of the approximations Eq. (\ref{app}) in this relativistic scenario
will be checked.
To this aim, the LF approach has been chosen 
due to its nice properties, in particular the fact that LF boosts
and plus component of the momenta ($a^+ = a_0 + a_z$) are kinematical
operators. The Fourier- transform of the dPDF 
\vskip -0.83cm
\begin{eqnarray} 
\label{ft}
F_{ij}^{\lambda_1,\lambda_2}(x_1,x_2,{\vec k}_\perp) 
= \int d \vec z_\perp \, e^{i \vec z_\perp \cdot \vec k_\perp} 
F_{ij}^{\lambda_1,\lambda_2}(x_1,x_2,{\vec z}_\perp)~,
\end{eqnarray}
will be analized.
In the above equation,
the dPDF is introduced for two
quarks of flavors $i$ and $j$ and helicities $\lambda_{i(j)}$, respectively.
The full procedure  of the calculations of the dPDF in the LF approach,
starting from the formal definition  of this quantity,
and thanks to a proper extension of the proceeding presented in Ref.
\cite{50a,51a}, developed in that case for the GPDs
calculations,
is shown in details in Ref. \cite{nostro}. Here the main steps
are summarized.
In particular, one can start from the expression of a light-cone
correlator (see, \eg, Ref. \cite{5a}), written in terms
of the proton state and of LF quantized fields of the
interacting quarks. In order to find a general expression of the
dPDF in the valence region, use has been made of the so
called ``LF wave function'' (LFWF) representation \cite{47a} to
describe the proton state. In this formalism the latter quantity
is written as a sum over partonic Fock states with all the  correct
normalizations  preserved. In the LFWF representation,
only the first term of this
summation, representing the contribution of the valence quarks,
has been taken into account. 
A crucial
point of the procedure is the possibility of describing the LF
proton state starting from the Instant Form (canonical) one,
where most quark models are developed. To this aim the following relation
between one particle LF state, $ | \vec k, \lambda \rangle_{[l]}$, and the
corresponding IF  one, $|\vec k, \lambda' \rangle_{[i]}$,  has been used:
\vskip -3mm
\begin{eqnarray}
\label{1pstate}
 | \vec k, \lambda \rangle_{[l]} = (2\pi)^{3/2} \sqrt{m^2+\vec k^2}~
\underset{\lambda \lambda'}{\sum} D^{1/2}_{\lambda \lambda'}(R_{cf}(\vec k))
|\vec k, \lambda' \rangle_{[i]}~,
\end{eqnarray}
\vskip -1mm
{ where 
the Melosh rotation, which allows to rotate the
canonical helicity, $\lambda$, into the LF spin, $\lambda'$, is introduced: }
\vskip -3mm
\begin{eqnarray}
D^{1/2}_{\mu
\lambda}(R_{cf}(\vec k)) =
\langle \mu  \left|~ \dfrac{m +x_iM_0 - i \vec 
\sigma_i
\cdot (\hat z \times \vec k_\perp)}{\sqrt{(m +x_i M_0)^2+\vec k_\perp^2}}
~\right| \lambda \rangle~.
\end{eqnarray}
\vskip -1mm
In the above equation,
$x_i = \dfrac{k_i^+}{P^+}$ is the longitudinal momentum fraction
carried by the $i$ parton, with $P^+$  the plus component of the proton
momentum, $M_0 = {\sum_i} \sqrt{m^2+ \vec k_i^2} $  the total
free energy mass of the partonic system,  ${ k_{iz} = -({m^2 +
k^2_{i \perp}-k^{+2}_i})/(2k^+_i)}$ , $\mu$ and $\lambda$ 
generic canonical spins.
Actually, we are interested in the proton, a composite system, and the
validity of Eq. (\ref{1pstate}), supposed for free states, is questionable.
Nevertheless, in the following, we will use a relativistic mass equation
built in accord with the Bakamjian-Thomas Construction of the Poincar\'e
generators, and Eq. (\ref{1pstate}) can be used (see Ref. \cite{46a}).
Using a lengthy but straightforward procedure,
a final expression of the dPDF is obtained. It reads \cite{nostro}:
\vskip -3mm
\begin{eqnarray}
\label{main}
F_{q_1 q_2}^{\lambda_1,\lambda_2}(x_1, x_2, \vec k_\perp) 
& = & 
3(\sqrt{3})^3 \int
\left[
\underset{i=1}{\overset{3}\prod} d \vec k_i 
\underset{\lambda_i^f \tau_i} {\sum}
\right]
\delta \left(
\underset{i=1}{\overset{3}\sum} \vec k_i 
\right) 
\Psi^* \left(\vec k_1 +
\dfrac{\vec k_\perp}{2}, \vec k_2 -
\dfrac{\vec k_\perp}{2},\vec k_3
; \{\lambda_i^f, \tau_i \}
\right) \\
\nonumber
& \times & \widehat{P}_{q_1}(1)\widehat{P}_{q_2}
(2)\widehat { P } _ { \lambda_1 } (1)\widehat { P } _ { \lambda_2 } (2)
\, \Psi \left(\vec k_1 -
\dfrac{\vec k_\perp}{2}, \vec k_2 +
\dfrac{\vec k_\perp}{2}, \vec k_3 
; \{\lambda_i^f, \tau_i \}
\right)
\\
\nonumber
& \times & \delta \left(x_1 
-\dfrac{k_1^+}{P^+}
\right) \delta \left(x_2 -\dfrac{k_2^+}{P^+}
\right)~.
\end{eqnarray}

\begin{figure}[t]
\begin{minipage}[t] {70 mm}
\vspace{7.0cm}
\includegraphics{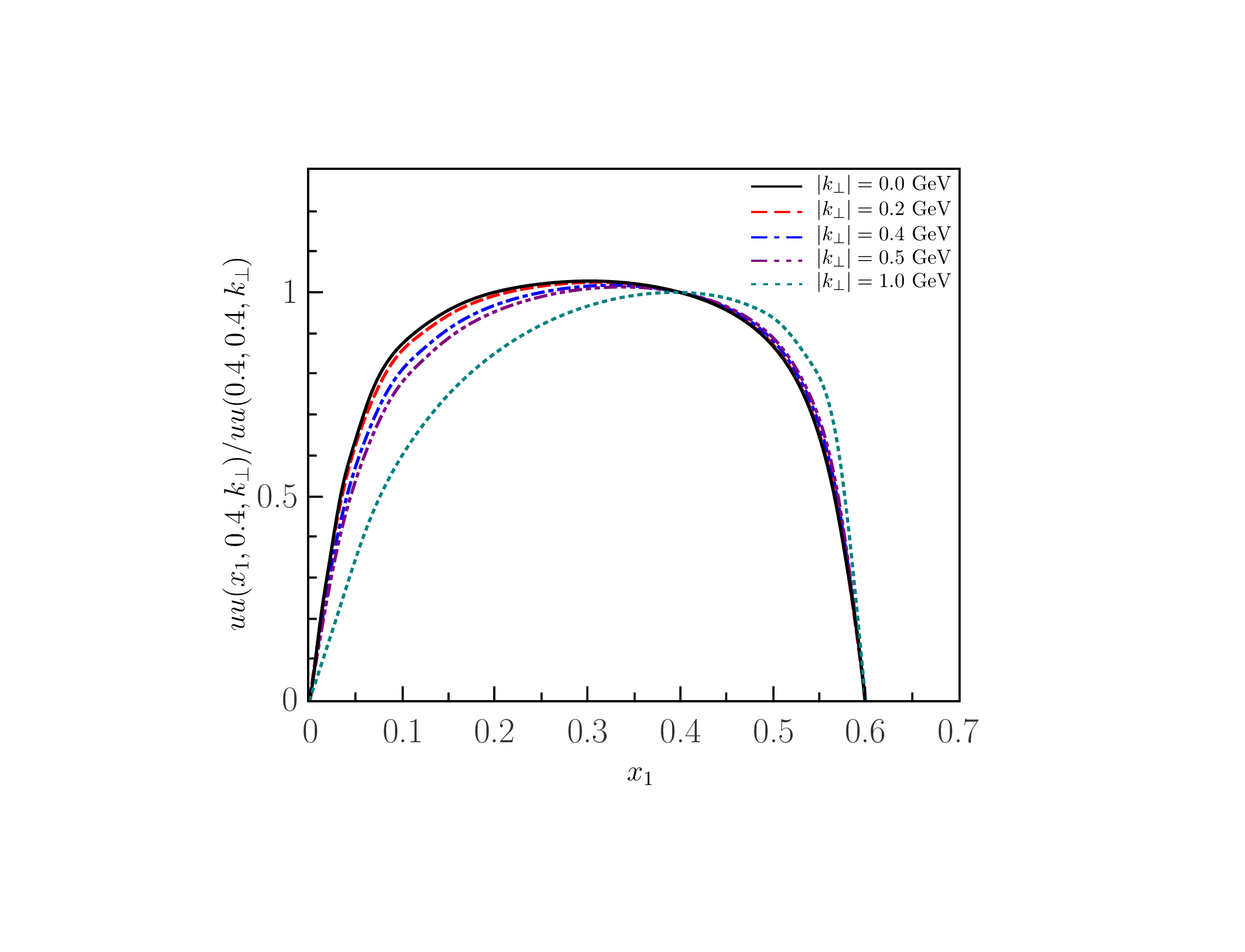}
\vskip -1.2cm
\caption{ \footnotesize The ratio $r_1$, Eq. (\ref{ratio12}), for
five values of $k_\perp$.}
\end{minipage}
\hspace{\fill}
\begin{minipage}[t] {70 mm}
\vspace{7.0cm}
\includegraphics{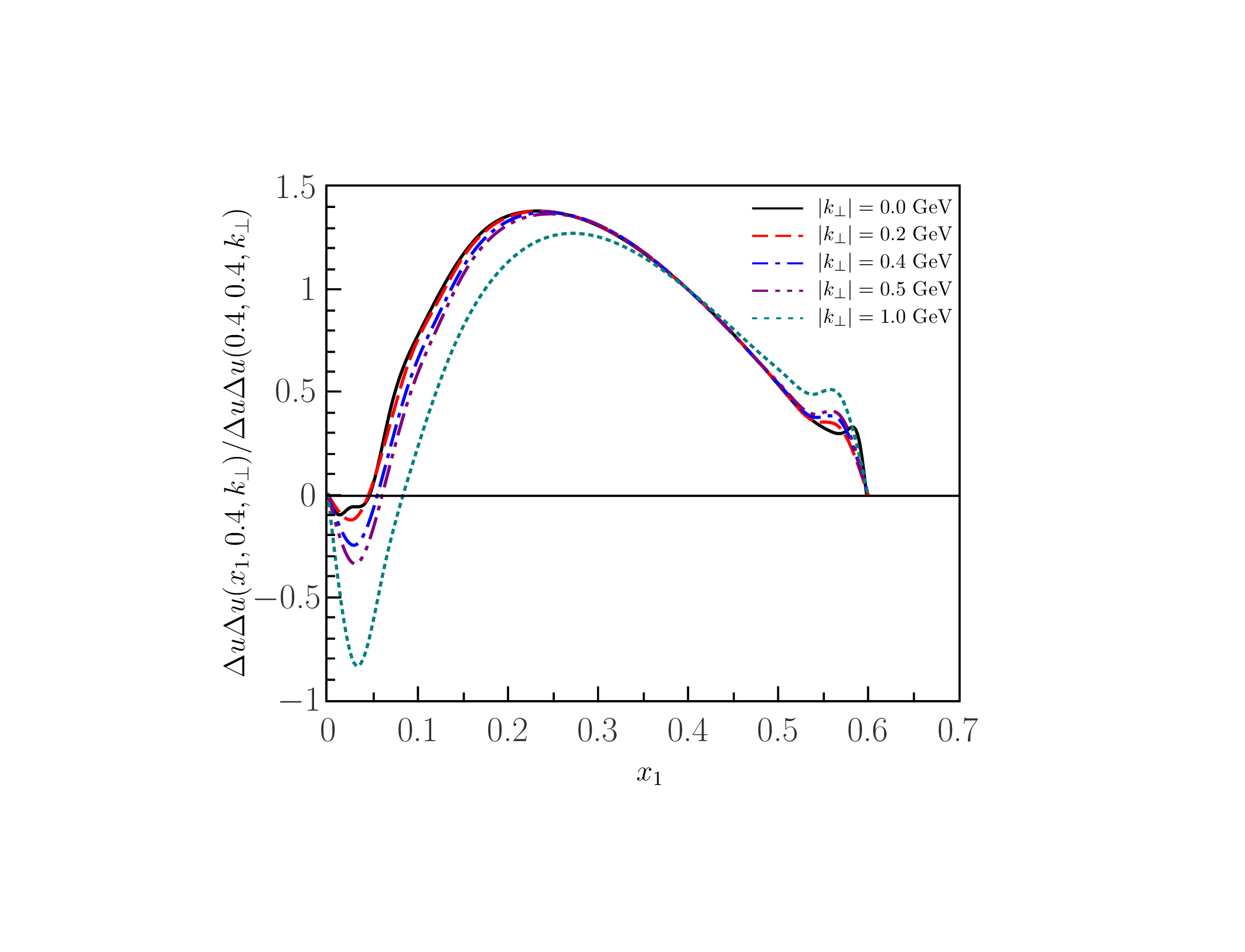}
\vskip -1.2cm
\caption{\footnotesize The ratio $r_2$, Eq. (\ref{ratio12}), 
for five values of $k_\perp$.}
\end{minipage}
\end{figure}

The canonical proton wave function $\psi^{[c]}$ is embedded in
the function
$\Psi$ here above, which can be written as follows:
\vskip -2mm
\begin{eqnarray}
\label{psiint}
\Psi(\vec k_1,\vec k_2,\vec k_2; \lbrace \lambda_i^f, \tau_i \rbrace) =
\underset{i=1}{\overset{3} \prod} \left [
\underset{\lambda_i^c}\sum D^{*1/2}_{\lambda_i^c
\lambda_i^f}(R_{cf}(\vec k_i))
\right ] \psi^{[c]}(\{\vec k_i, \lambda_i^c, \tau_i   \} )~,
\end{eqnarray}
\vskip -1mm
where here $\lambda_i^c$, $\tau_i$ are the canonical partonic helicity and
isospin respectively and the short notation $\{ \alpha_i \}$ instead of
$\alpha_1, \alpha_2, \alpha_3$ is introduced. Moreover isospin and spin
projection operators are introduced in order to access  dynamical
correlations for the unpolarized and longitudinal polarized $i$ quark of a
given flavor:
\vskip -2mm
\begin{eqnarray}
\label{proj}
\hat P_{u(d)}(i) = \dfrac{ 1 \pm \tau_3(i)}{2}~, ~
\hat P_{\lambda_k}(i)  =  \dfrac{ 1 + \lambda_k \sigma_3(i)}{2}~.
\end{eqnarray}
\vskip -1mm
A crucial point of this analysis, as already pointed out, is that now the
plus components of the momenta are totally kinematical so that one finds:
\vskip -.5cm
\begin{eqnarray}
\label{pp}
P^+= \overset{3}{ \underset{i}\sum} k_i^+  = M_0~,
\end{eqnarray}
\vskip -3mm
in the
intrinsic frame where $\vec k_1 + \vec k_2+ \vec k_3=0$~.
{ Thanks to the relation Eq. (\ref{pp})} the delta function, defining the
longitudinal
momentum fraction carried by the parton in Eq. (\ref{main}), can be
properly solved without any additional approximation, at variance with
what happens in the instant form 
calculation in Ref. \cite{36a}. As a direct consequence, the bad support
problem does not show up.
The following distributions,
different from zero for an unpolarized proton, have been calculated:
\vskip -2mm
\begin{eqnarray}
 \label{unp}
uu(x_1,x_2, k_\perp) = \underset{i,j = \uparrow,\downarrow}\sum
u_iu_j(x_1,x_2, k_\perp),~~ \Delta u \Delta u(x_1,x_2, k_\perp) =\underset{i,j =
\uparrow,\downarrow}\sum (-1)^{i+j+1}
u_iu_j(x_1,x_2, k_\perp)
\end{eqnarray}
\vskip -2mm
In order to calculate now the dPDFs, in
particular to check whether the approximation, Eq. (\ref{app}), holds, a
proper CQM has to be chosen. To this aim, in order to have a fully
consistent procedure, a relativistic model has been used, in particular the
one described in Ref. \cite{49a}, a hyper-central CQM. This
model provides a reasonable description of the light hadronic spectrum, it
has been used for the estimate of PDFs and GPDs in Refs. \cite{49a,
50a,51a,54a,55a} and,
for the present analysis, since no data are available for the
dPDF, it can be used as laboratory
to predict the most relevant features of dPDF.
In particular, in Figs. 1 and 2 the following ratios have been shown for
five values of $k_\perp$:
\begin{eqnarray}
\label{ratio12}
 r_1 = \dfrac{uu(x_1,0.4,k_\perp )}{uu(0.4,0.4,k_\perp )},~~r_2 =
\dfrac{\Delta u \Delta u(x_1,0.4,k_\perp )}{\Delta u \Delta
u(0.4,0.4,k_\perp )}~.
\end{eqnarray}
{ In particular,  the value  $x_2 = 0.4$ has been chosen for an easy
comparison with the results of Refs.
\cite{33a,36a}}.
 From these results it is clear that the factorization ansatz, in Eq.
(\ref{app}), is  violated,  as it was found already
in Refs. \cite{33a,36a}. { As one can notice,  $r_1$ and
$r_2$  depend on $k_\perp$ so that a factorized form of the dPDFs for
the
$k_\perp$ dependence is not  supported by  this approach}. One can easily
realize, in fact, that a factorized expression for the dPDF would yield
$k_\perp$-independent $r_1$ and $r_2$. It
is also important to notice that the amount of the violation is directly
related to the  Melosh rotation contributions, a model independent
relativistic effect.
In Fig. 3 the following ratios
\vskip -3mm
\begin{eqnarray}
\label{ratio34}
 r_3 = \dfrac{2 uu(x_1,x_2, k_\perp=0)}{u(x_1)u(x_2)}~,~
 r_4 = \dfrac{C \Delta u \Delta u(x_1,x_2, k_\perp=0)}{\Delta u(x_1)
\Delta u(x_2)}~,
\end{eqnarray}
\vskip -1mm
where:
\begin{eqnarray}
 C = \dfrac{ [\int dx \Delta u(x)]^2  }{\int d x_1 d x_2 \Delta u
\Delta u(x_1,x_2,k_\perp =0)}~,
\end{eqnarray}
involving the
standard PDFs in the denominators, calculated within the same CQM, 
have been presented.
\begin{figure}[t]
\vspace{15.0cm}
\includegraphics{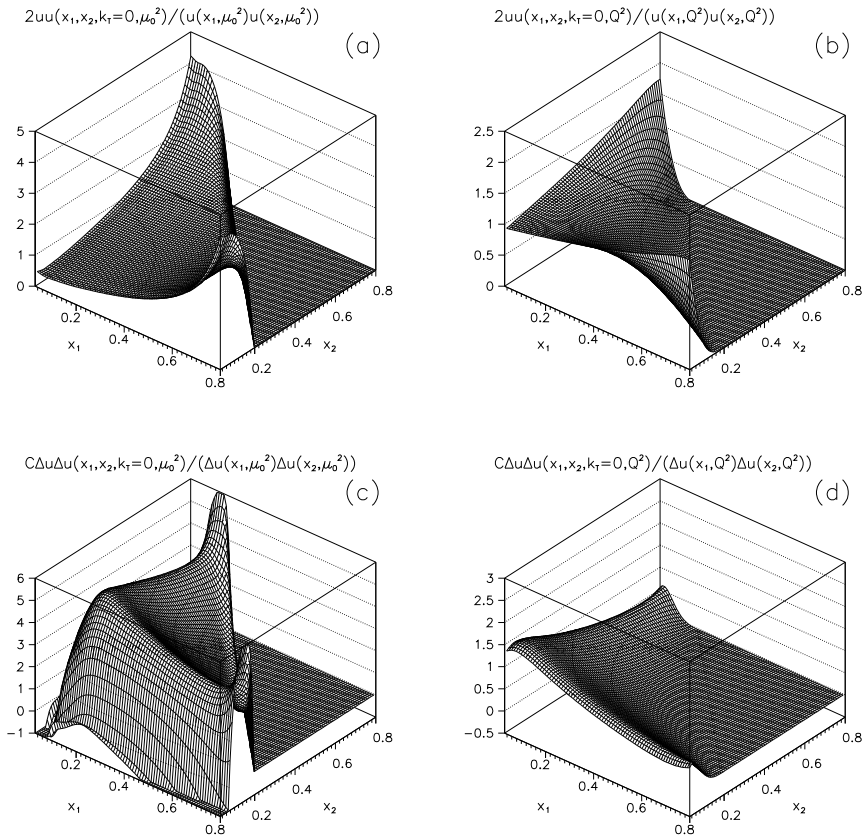}
\vskip -3cm
\caption{ \footnotesize a) The ratio $r_3$, Eq. (\ref{ratio34}), at the
hadronic scale;
b) the same quantity at a scale $Q^2= 10$ GeV$^2$;
c) the ratio $r_4$, Eq. (\ref{ratio34}), at the hadronic scale $\mu_0$;
d) this last quantity at a scale $Q^2= 10$ GeV$^2$.
The vertical scale of panels (b) and (d) is reduced by a factor of 2
with respect to panels (a) and (c), respectively.} 
\end{figure}
In Figs. 3(a), 3(c) these quantities are shown  at
the hadronic scale obtaining that a factorized approximation of the dPDF,
in terms of the single PDFs, is not favored in the valence quark region.
Notice that the factors $2$ and $C$, in Eq. (\ref{ratio34}), are 
inserted in order to have these ratios equal to $1$ in the kinematical
region
where the approximation, Eq. (\ref{app}), is valid. Moreover, in Ref.
\cite{nostro} the pQCD evolution at Leading-Order of the calculated dPDFs
has been also presented. In particular, for the moment being,  the evolution
of the dPDF has been addressed for  $k_\perp =0$, taking the same
scale for the two acting partons and analyzing only the valence quark
contributions. In this case  the evolution equations are obtained as
a direct generalization of the well known DGLAP ones, see Refs.
\cite{23a,24a}. Since only the  non singlet case has
been studied here, one needs to solve  the homogeneous part of the
generalized DGLAP equations by using the Mellin transformations of the
dPDFs, see Ref. \cite{nostro} for details. 
If we use their simple ansatz as an input to our calculation, our code
reproduce the results of the authors  of Ref. \cite{30a}. 
Once the dPDFs have been evaluated
at a generic high energy scale, \eg, $Q^2 = 10$ GeV$^2$, the ratios $r_3$
and $r_4$ have been shown again in Figs. 3(b), 3(d). The most important
results of this analysis are that, for small values of $x$, where data
are taken, \eg, at the LHC,  $r_3 \sim 1$, which means that, in the
unpolarized case, dynamical correlations are suppressed after the
evolution. Nevertheless, by looking at $r_4$, in Fig 3(d), it is found that
double spin
correlations still contribute. \\ 
\vskip 1mm
{\bf 3 Conclusions} \\
\vskip 0.5mm
In this work, dPDFs contributing to the DPS cross section have been
calculated by means of a LF CQM. The main achievement is the fully Poincar\'e 
covariance of the description, which
allows to restore the expected symmetries, and the vanishing of the dPDF in
the forbidden kinematical region, $x_1+x_2>1$. 
In the analysis of the dPDFs at the hadronic scale we
found that the approximations of these quantities with a
complete factorized ansatz, in the $x_1-x_2$ and $(x_1,x_2)-k_\perp$
dependences, is violated, in agreement with  previous results  
\cite{33a,36a}. Moreover, a pQCD analysis of the dPDFs, necessary to
evaluate these quantities at higher energy scales with respect to the
hadronic one where the CQM predictions are valid, has been
also performed. For the unpolarized quarks case the dynamical
correlations are suppressed in the small $x$ region, while double
spin correlations are found to be still important. Further analysis, 
including into the
scheme non perturbative sea quarks and gluons and the evolution of the
singlet sector, important to describe the dPDF at low $x$,
are in progress, as well as the study of correlations in $pA$ scattering, 
along the line of Ref. \cite{25}.

\end{document}